\documentclass[twocolumn,showpacs,amsmath,amssymb,10pt]{revtex4}
\usepackage{amsfonts}
\usepackage{bm}
\begin{document}
\title{\textbf{Wigner function and Schr\"{o}dinger equation in
phase space representation}} \author{Dariusz
Chru\'sci\'nski\thanks{email: darch@phys.uni.torun.pl} and
Krzysztof M\l odawski\thanks{email: simson@phys.uni.torun.pl}}
\affiliation{Institute of Physics, Nicolaus Copernicus University,\\
Grudzi\c{a}dzka 5/7, 87--100 Toru\'n, Poland}

\begin{abstract}
We discuss a family of quasi-distributions (s-ordered Wigner
functions of Agarwal and Wolf) and its connection to the so called
phase space representation of the Schr\"odinger equation. It turns
out that although Wigner functions satisfy the Schr\"odinger
equation in phase space they have completely different
interpretation.

\end{abstract}

\maketitle

\section{INTRODUCTION}\label{A0}

Since the pioneering work of Wigner \cite{wig}, generalized
phase-space techniques have found useful applications in various
branches of physics \cite{app1,app2,Balazs,rep}. The main idea of
this approach is to represent the density operator $\hat{\rho}$ as
a function (quasi-distribution) over the classical phase space
$(q,p)$. This function fully characterized the quantum state and
enables one to express the quantum-mechanical expectations as
averages of classical observables over the classical phase space.
Moreover, it is the Wigner function which is directly related to
the measurement. Then  quantum tomographic methods
\cite{leon,Dariano} enable one to reconstruct the quantum state
from the experimental data. Recently, the Wigner function was also
applied to study quantum entanglement and related issues for
continuous systems (see e.g. \cite{ent1,ent2,ent3}).

The Wigner function \cite{wig} is only one particular example of
such a quasi-distribution. Especially important role (e.g. in
quantum optics) is played by the family of functions introduced by
Cahill and Glauber \cite{Cahill} containing as the Wigner
function, the Glauber-Sudarshan $P$-function \cite{P1,P2}, and the
Husimi $Q$-function \cite{Q}. In this paper we analyze another
lesser-known family introduced by Agarwal and Wolf \cite{Agarwal}.
Actually, all these quasi-distributions correspond to the
particular quantization procedure, that is, different ordering of
$\hat{q}$ and $\hat{p}$, or, equivalently,  different ordering of
creation $\hat{a}^\dag$ and $\hat{a}$ annihilation operators,
respectively.

The procedure of representing quantum states by
quasi-distributions in phase space is closely related to the phase
space formulation of quantum mechanics based on the noncommutative
product  known as Moyal product \cite{Groenewold,Moyal} or more
generally as a star-product \cite{Flato} (see \cite{amer,Zachos}
for the compact formulation of the standard quantum mechanics in
terms of the Moyal product).

There is another phase space representation of quantum mechanics
based on the works of Torres-Vega and Frederic
\cite{Torres1,Torres2} (see also \cite{har}). In this approach the
(pure) quantum state is represented by the wave function
$\psi(\Gamma)$, where $\Gamma$ represents a point in phase space.
It turns out that $\psi(\Gamma)$ satisfies so called
``Schr\"odinger equation in phase space''. The quantity
$|\psi(\Gamma)|^2$ is, therefore, treated as a probability
distribution in phase space.
 This procedure was applied to study simple quantum systems \cite{t1,t2,scripta}.
In a recent paper Li, Wei and L\"{u} \cite{li} found a general
method of solving ``Schr\"odinger equation in phase space''.

The aim of the present paper is to relate the standard phase space
approach based on quasi-distributions functions to that of
Torres-Vega and Frederic (TF). We show that one can easily produce
the whole family of ``Schr\"odinger equation in phase space''
which is closely related to the family of $s$-ordered Wigner
functions $W_s$ of Agarwal and Wolf \cite{Agarwal}, that is, $W_s$
are particular solutions of this family of equations. Now,
according to the standard approach $W_s$ defines the
quasi-distribution in phase space, whereas the TF approach implies
that $|\psi_s|^2$, where $\psi_s = 2\pi\hbar\, W_s$,  is a (true)
probability distribution, i.e. $|\psi_s|^2\geq 0$ and $\int
|\psi_s|^2\, d\Gamma=1$. It should be stressed, that
``Schr\"odinger equation in phase space'' has an infinite number
of solutions. It is a price one pays for using $\psi(\Gamma)$
instead of $\varphi(q)$. Each particular solution gives rise to
the particular phase space representation of ordinary wave
function $\varphi(q)$ in position representation.

 The paper is organized as follows. In Sec. \ref{A1} we present
 general approach to phase space representation of the wave
function and following \cite{li} we discuss the general solution
for ``Schr\"odinger equation in phase space''. Section \ref{A2}
introduces the whole $s$-family of equations together with the
family of solutions. Then, after recalling the formulae for
star-products in Section \ref{A3}  we show that $s$-ordered Wigner
functions do solve the family of Schr\"odinger equations. We end
with some conclusions in Section \ref{A5}.

\section{SCHR\"{O}DINGER EQUATION IN PHASE SPACE}\label{A1}

There is no a unique way to represent a quantum state as a wave
function $\psi=\psi(\Gamma)$, where $\Gamma$ represents a point in
a classical phase $(q,p)$. In the standard approach one usually
uses a coordinate  $\varphi(q)$ or momentum
$\widetilde{\varphi}(p)$ representations, respectively. To pass
from $\varphi(q)$ to $\psi(\Gamma)$ one has to invent an integral
transformation
\begin{equation}\label{Kernel}
    \psi(\Gamma) = \int K(\Gamma;q') \, \varphi(q')\, dq'\ ,
\end{equation}
where $K(\Gamma;q')$ denotes the integral kernel. Functions
$\psi(\Gamma)$ defined by the above formula form a proper subspace
${\cal H}_K$ of the Hilbert space $L^2( \mathbb{R}^2)$. The
unitarity of transformation $L^2( \mathbb{R}) \ni \varphi(q)
\longrightarrow \psi(\Gamma) \in {\cal H}_K$ requires
\begin{equation}\label{KK}
    \int \overline{K}(\Gamma;q')\, K(\Gamma;q'')\, d\Gamma =
    \delta(q'-q'')\ ,
\end{equation}
where $d\Gamma$ denotes a measure on the phase space. Clearly,
there is a huge freedom in choosing $K$. Performing the following
``gauge transformation"
\begin{equation}\label{}
K(\Gamma;q) \longrightarrow e^{if(\Gamma)}\, K(\Gamma;q)\ ,
\end{equation}
with $f(\Gamma)$ being an arbitrary real function, one obtains a
new kernel still satisfying (\ref{KK}).

In the literature there are several well known examples of such a
transform. Perhaps the most famous is the Bargmann (or
Bargmann-Segal) transform defined by \cite{Bargmann,Segal}
\begin{equation}\label{}
    K_{\rm B}(z;q) = \pi^{-1/4}\, \exp\left\{ - \frac 12 (z^2 + q^2) +
    \sqrt{2}zq \right\} \ ,
\end{equation}
where $z$ is a complex number, i.e. one uses $ \mathbb{R}^2 \cong
\mathbb{C}$. The corresponding  space ${\cal H}_{\rm B}$ of entire
functions $\psi(z)$  equipped with the following inner product
\begin{equation}\label{}
    \langle \psi_1|\psi_2\rangle_{\rm B} = \int \psi_1^*(z)\,
    \psi_ 2(z)\, d\mu(z)\ ,
\end{equation}
where $d\mu(z) = \pi^{-1}e^{-|z|^2}d^2z$, is known as the
Bargmann-Segal representation of the Hilbert space.

A closely related kernel is connected to the coherent states
representation \cite{Klauder}
\begin{equation}\label{K-CS}
 K_{\rm CS}(q,p;q') =  \langle \Gamma|q'\rangle\ ,
\end{equation}
where
\begin{eqnarray}\label{}
\langle \Gamma|q'\rangle\   =  &&\nonumber\\ && \hspace{-2cm}
(\lambda^2\pi)^{-1/4}    \exp\left\{ - \frac{(q' -
    q)^2}{2\lambda^2}
    - \frac{i p}{\hbar}(q'-q) \right\} \ ,
\end{eqnarray}
with $|\Gamma\rangle$ denoting the standard Glauber coherent state
corresponding to
\begin{eqnarray}\label{Gamma}
\Gamma = (\lambda^{-1}q+i\lambda\hbar^{-1}p)/\sqrt{2}\ .
\end{eqnarray}
 The
parameter $\lambda$ is a natural length scale defined by the mass
and frequency of the oscillator, i.e. $\lambda =
\sqrt{\hbar/\mu\omega}$.
 The corresponding
Hilbert space ${\cal H}_{\rm CS}$ carries the following inner
product
\begin{equation}\label{}
\langle \psi_1|\psi_2\rangle_{\rm CS} =  \int \psi_1^*(\Gamma)\,
    \psi_ 2(\Gamma)\, d\Gamma\ ,
\end{equation}
with
\begin{equation}\label{d-Gamma}
d\Gamma= \frac{dqdp}{{2\pi\hbar}}\ .
\end{equation}
From now on we shall use the dimensionless convention
(\ref{d-Gamma}) for $d\Gamma$. Note, that in this convention
$\psi(\Gamma)$ is dimensionless, whereas the kernel $K(\Gamma;q')$
has the same dimension as the wave function in the position
representation $\varphi(q)$.

 Actually, the formula  (\ref{K-CS}) was a
starting point in constructing phase space representation of
quantum mechanics of Torres-Vega and Frederick
\cite{Torres1,Torres2}. They showed that if $\varphi(q)$ satisfies
the standard Schr\"odinger equation
\begin{equation}\label{}
    i\hbar \frac{\partial}{\partial t}\,\varphi(q,t) = \left[
    -\frac{\hbar^2}{2m} \, \frac{\partial^2}{\partial q^2} + V(q)
    \right] \varphi(q,t)\ ,
\end{equation}
then $\psi(\Gamma)$ obtained from $\varphi(q)$ via $K_{\rm
CS}(\Gamma;q)$ satisfies the following ``Schr\"odinger equation in
phase space''
\begin{equation}\label{}
    i\hbar \frac{\partial}{\partial t}\,\psi(\Gamma,t) = \left[
    -\frac{\hbar^2}{2m} \, \frac{\partial^2}{\partial q^2} +
    V\left(q+ i\hbar\frac{\partial}{\partial p}\right)
    \right] \psi(\Gamma,t)\ .
\end{equation}
Actually, performing the gauge transformation $\psi'(q,p) =
e^{-ipq/2\hbar}\psi(q,p)$ one finds more symmetric formula
\begin{equation}\label{}
i\hbar \frac{\partial}{\partial t}\,\psi'(\Gamma,t) = \left[
\frac{1}{2m} \hat{P}^2 + V(\hat{Q}) \right] \psi'(\Gamma,t)\ ,
\end{equation}
where
\begin{eqnarray}\nonumber
\hat{Q} &=& \frac{q}{2}+i\hbar\frac{\partial}{\partial p}\ ,
  \\\label{A}
\hat{P}&=&\frac{p}{2}-i\hbar\frac{\partial}{\partial
q}\ ,
\end{eqnarray}
satisfy  $[\hat{Q},\hat{P}]=i\hbar$ and, therefore, they define
phase space representation of position and momentum. This
particular representation corresponds to the gauge transformed
coherent states kernel $e^{-ipq/2\hbar}K_{\rm CS}(q,p;q')$.

Recently, the following stationary Schr\"odinger equation
\begin{equation}\label{B}
\left[\frac{1}{2m}\left(\frac{p}{2}-i\hbar\frac{\partial}{\partial
q}\right)^2 +V\left(\frac{q}{2}+i\hbar\frac{\partial}{\partial
p}\right)\right]\psi(\Gamma)=E\psi(\Gamma)\ ,
\end{equation}
was postulated in \cite{li}. Now, $\psi(\Gamma)$ denotes an
arbitrary phase space representation, that is, the integral kernel
$K(\Gamma;q)$ in (\ref{Kernel}) is not specified. The general
solution of (\ref{B}) reads as follows (eq. (11) in \cite{li})
\begin{equation}\label{x1}
\psi(\Gamma)=e^{-iqp/2\hbar}\int
g(y)\varphi(q+y)e^{-\frac{i}{\hbar}py}\, dy\ ,
\end{equation}
where $g(y)$ is an arbitrary nonzero function and $\varphi(q)$ is
the eigenfunction of the Schr\"{o}dinger equation in coordinate
representation corresponding to the eigenvalue $E$. Note, that the
function $g(y)$ uniquely defines an integral kernel $K_g$ by
\begin{equation}\label{K-g}
    K_g(q,p;q')  = e^{-ip(q' - q/2)/\hbar}\, g(q'-q)\ .
\end{equation}
Note, that
\begin{eqnarray}\label{}
\lefteqn{
\int \overline{K}(\Gamma;q')\, K(\Gamma;q'')\, d\Gamma} \nonumber\\
&&  =    \delta(q'-q'')  \int {g}^*(q'-q)g(q''-q)\, dq\ ,
\end{eqnarray}
and hence unitarity condition (\ref{KK}) implies
\begin{equation}\label{g2}
\int |g(y)|^2\, dy = 1\ .
\end{equation}
 In particular, the following Gaussian
\begin{equation}\label{g-ex}
    g(y) = (\pi \lambda^2 )^{-1/4}\, e^{-y^2/2\lambda^2}\
    ,
\end{equation}
does satisfy (\ref{g2}) and one finds for the corresponding
$K_g(q,p;q') =e^{-ipq/2\hbar}K_{\rm CS}(q,p;q')$.

\section{A FAMILY OF THE SCHR\"{O}DINGER EQUATION IN PHASE
SPACE}\label{A2}

Let us observe that the representation (\ref{A}) may be
generalized to the whole family of representations. It is
convenient to scale phase space variables $(q,p) \rightarrow
(2q,2p)$ and to introduce
\begin{eqnarray} \label{As}\nonumber
\hat{Q}_s &=& q + (1-s)\frac{i\hbar}2\frac{\partial}{\partial p}\ ,\\
\label{x2}\hat{P}_s&=&  p
-(1+s)\frac{i\hbar}2\frac{\partial}{\partial q },
\end{eqnarray}
with $s \in \mathbb{R}$. One can easily verify that
$[\hat{Q}_s,\hat{P}_s]=i\hbar$ for  $s\neq\pm 1$. Note however,
that $\lim_{s\rightarrow \pm 1}[\hat{Q}_s,\hat{P}_s]=i\hbar$.
Therefore, the values $s=\pm 1$ will be understood as appropriate
limits.

In analogy to (\ref{B})  let us postulate the following family of
Schr\"{o}dinger equations
\begin{eqnarray}\label{x3}
\Bigg[ \!\!\! && \left. \!\!\!\frac{1}{2m}\left(
p-(1+s)\frac{i\hbar}{2}\frac{\partial}{\partial q}\right)^2+
V\left(q + (1-s)\frac{i\hbar}2\frac{\partial}{\partial
p}\right)\right] \nonumber\\
&& \times\, \psi_s(\Gamma)=E\psi_s(\Gamma)\ .
\end{eqnarray}
 To solve this equation assume that
\begin{equation}\label{x4}
\psi_s(q,p)=\exp\left\{\frac{-2i}{\hbar}\frac{pq}{1+s}\right\}\phi_s(q,p)\
.
\end{equation}
One obtains the following equation for $\phi_s$:
\begin{eqnarray}\label{x5}
\Bigg[ \!\!\!&-&\!\!\!
\frac{\hbar}{2m}\frac{(1+s)^2}{4}\frac{\partial^2}{\partial q^2}+
V\left(\frac{2q}{s+1}+(1-s)\frac{i\hbar}{2}\frac{\partial}{\partial
p}\right)\Bigg]  \nonumber \\ &\times& \phi_s(q,p) =E\phi_s(q,p)\
.
\end{eqnarray}
Now we expand  the potential $V$ as a Taylor's series about
$(i\hbar(1-s)/2)\partial/\partial p$ for given $q$ and use the
partial Fourier transform
\begin{equation}\label{x6}
\phi_s(q,p)=\int
\chi_s(q,y)e^{-ipy/\hbar}\ .
\end{equation}
Further, multiplying both side of (\ref{x5}) by
$\exp\{-ipy'/\hbar\}$ and integrating over $p$, one obtains
\begin{eqnarray}\label{x7}
\Bigg[ \!\!\!&-&\!\!\!
\frac{\hbar^2}{2m}\frac{(1+s)^2}{4}\frac{\partial^2}{\partial
q^2}+ V\left(\frac{2}{s+1}q+\frac{1-s}{2}y\right)\Bigg]\chi_s(q,y)
\nonumber \\ && = E\chi_s(q,y)\ ,
\end{eqnarray}
which defines the standard Schr\"odinger equation in the
$\xi$--representation
\begin{equation}\label{xi}
    \xi = \frac{2}{s+1}q+\frac{1-s}{2}y\ .
\end{equation}
Therefore, the general solution  of (\ref{x7}) reads as follows
\begin{equation}\label{x8}
\chi_s(q,y)=g_s(y)\varphi(\xi)\ ,
\end{equation}
where $g_s=g_s(y)$ is an arbitrary nonzero function and
$\varphi(\xi)$ satisfies
\begin{equation}\label{}
     \left[
    -\frac{\hbar^2}{2m} \, \frac{\partial^2}{\partial \xi^2} + V(\xi)
    \right] \varphi(\xi) = E \varphi(\xi)\ .
\end{equation}
Finally, the general solution of (\ref{x4}) has the following form
\begin{eqnarray}\label{x9}
\psi_s(q,p)&=&\exp\left\{-\frac{2i}{\hbar}\frac{pq}{s+1}\right\}\int
dy\, e^{-{i}py/\hbar} \nonumber \\ &\times& g_s(y)
\varphi\left(\frac{2}{s+1}q+\frac{1-s}{2}y \right)\ .
\end{eqnarray}
Clearly, for each $g_s(y)$ it defines a family of kernels $K^s_g$
\begin{equation}\label{}
\psi_s(q,p) = \int K^s_g(q,p;q')\, \varphi(q')\, dq'\ ,
\end{equation}
given by
\begin{eqnarray}\label{Ksg}
    K^s_g(q,p;q') &=& \frac{2}{1-s}\, g_s\left( \frac{2q'}{1-s} -
    \frac{4q}{1-s^2} \right)\nonumber \\ &\times&  \exp\left(-
    \frac{2i}{1-s}\frac{p(q'-q)}{\hbar}\right) \ .
\end{eqnarray}
Again, the requirement of unitarity  (\ref{KK}) implies the
following condition for the function $g_s$:
\begin{equation}\label{g2-s}
    \int |g_s(y)|^2\, dy = \frac{2}{|1+s|}\ .
\end{equation}
Note that for $s=1$ the formula for the kernel considerably
simplifies
\begin{equation}\label{}
    K_g^{s=1}(q,p;q') = e^{-ipq/\hbar}\,
    \widetilde{g}(p)\delta(q-q')\ ,
\end{equation}
where $ \widetilde{g}$ stands for the Fourier transform of $g$. In
this case the ave function $\varphi(q)$ has the following phase
space representation
\begin{equation}\label{}
    \psi(q,p) = e^{-ipq/\hbar}\,
    \widetilde{g}(p)\varphi(q)\ .
\end{equation}

\section{STAR PRODUCT}\label{A3}

Now we show that the family of equations (\ref{x3}) is closely
related to the family of quasidistributions function in phase
space.

To describe all quantum phenomena in phase space, we have to
determine a relationship between operators and functions on the
classical phase space. This correspondence is of course not
unique. The most famous is based on the Wigner--Weyl transform
${\cal F}_{\rm WW}$: if $f(q,p)$ is a phase space function then
one defines the corresponding operator $\hat{F}$
\[  \hat{F} = {\cal F}_{\rm WW}(f) \ , \]
by
\begin{equation} \label{WW}
\hat{F}=\int
d\sigma \int
d\tau\, \tilde{f}(\sigma,\tau)e^{i(\sigma\hat{q} +\tau\hat{p})}\ ,
\end{equation}
where $\tilde{f}$ denotes the Fourier transform of $f$
\begin{equation}\label{x10}
\tilde{f}(\tau,\sigma)=\frac{1}{2\pi}\,
\int
d\sigma\int
d\tau\, f(q,p)e^{-i(\sigma q +\tau p)}\ .
\end{equation}
The inverse transform, i.e. $\hat{F} \rightarrow f$, is defined as
follows
\begin{equation}\label{cztery}
f(q,p)=\int
dy\, \left\langle q-\frac{1}{2}y\left\vert\hat{F}\right\vert
q+\frac{1}{2}y\right\rangle e^{{i}py/{\hbar}}\ .
\end{equation}
If $\hat{F}$ corresponds to a density operator $\hat{\rho}$ then
its inverse Wigner-Weyl transform recovers celebrated Wigner
function  $W = {\cal F}^{-1}_{\rm WW}(\hat{\rho}) $
\begin{equation}\label{x12}
W(q,p)=\frac{1}{\pi\hbar}\int
dy\, \langle q-y\vert\hat{\rho}\vert q+y\rangle
e^{{2i}py/{\hbar}}\ .
\end{equation}

Now, the noncommutative multiplication   of  operators introduces
the following noncommutative multiplication of functions
\begin{equation}\label{AB}
    \hat{F}_1 \cdot \hat{F}_2 = {\cal F}_{\rm WW}(f_1 \star f_2)\ .
\end{equation}
The formula for the star-product `$\star$' was derived long ago by
Groenewold and Moyal \cite{Groenewold,Moyal}:
\begin{equation}\label{E}
f_1\star f_2 = f_1
\exp\left\{\frac{i\hbar}{2}\left(\overleftarrow{\partial}_q\overrightarrow{\partial}_p
-\overleftarrow{\partial}_p\overrightarrow{\partial}_q\right)\right\}f_2
\ ,
\end{equation}
where $\overleftarrow{\partial}$ ( $\overrightarrow{\partial}$ )
act on the left (right) side. The Moyal product is associative
\[  (f_1 \star f_2) \star f_3 = f_1 \star ( f_2 \star f_3) \ , \]
but it is noncommutative
\[  f_1 \star f_2 \neq f_2 \star f_1\ .  \]
Recall, that the operator $e^{a\partial_x}$ acts as a generator of
translations: $e^{a\partial_x}f(x) = f(x+a)$. Therefore, the
defining formula (\ref{E}) may be rewritten in the following form
\begin{equation}\label{}
    f_1(q,p) \star f_2(q,p) = f_1\left(q +
    \frac{i\hbar}{2}\partial_p\, ,  p -
    \frac{i\hbar}{2}\partial_q\right) f_2(q,p)\ .
\end{equation}
Now we can make crucial observation: equation (\ref{x3}) for $s=0$
has the following form
\begin{equation}\label{s=0}
    H \star \psi_{s=0} = E \psi_{s=0}\ ,
\end{equation}
where $H(q,p)$ is the classical Hamilton function.

It turns out that the similar structure may be established also
for $s\neq 0$. Let us introduce  the following family of
Wigner-Weyl transforms:
\[  \hat{F}_s = {\cal F}^s_{\rm WW}(f) \ , \]
by
\begin{equation} \label{WW-s}
\hat{F}_s
=\int
d\sigma\int
d\tau\, \tilde{f}(\sigma,\tau)e^{i(\sigma\hat{q}
+\tau\hat{p})}e^{-is\sigma\tau/2}\ .
\end{equation}
Clearly, for $s=0$ one recovers (\ref{WW}). This formula enables
one to introduce the family of star-products `$\star_s$'
\begin{equation}\label{AB-s}
    \hat{A}_s \cdot \hat{B}_s = {\cal F}^s_{\rm WW}(a \star_s b)\ ,
\end{equation}
which reduces to (\ref{AB}) for $s=0$. One easily finds
\begin{equation}\label{E-s}
a\star_s\, b = a
\exp\left\{\frac{i\hbar}{2}\left((1-s)\overleftarrow{\partial}_q\overrightarrow{\partial}_p
-(1+s)\overleftarrow{\partial}_p\overrightarrow{\partial}_q\right)\right\}b
\ ,
\end{equation}
which is equivalent to
\begin{eqnarray}\label{}
   \lefteqn{ a(q,p) \star_s\, b(q,p) =} \nonumber\\ && \hspace*{-.5cm} a\left(q +
    (1-s)\frac{i\hbar}{2}\partial_p\, ,  p -
    (1+s)\frac{i\hbar}{2}\partial_q\right) b(q,p)\ .
\end{eqnarray}
Therefore, the family of Schr\"odinger equations in phase space
(\ref{x3}) may be rewritten as follows:
\begin{equation}\label{star-psi-s}
    H \star_s\, \psi_{s} = E \psi_{s}\ .
\end{equation}
This shows that the family of equations (\ref{x3}) which is a
direct generalization of equations used in \cite{Torres1} and
\cite{li} is closely related to the noncommutative structure
induced by the family of star-products.

\section{WIGNER FUNCTION VS. PHASE SPACE WAVE FUNCTION}\label{A4}

Both Wigner function $W(q,p)$ and the wave function $\psi(\Gamma)$
are objects defined on the classical phase space. $\psi(\Gamma)$
satisfies (\ref{s=0})
\begin{equation}\label{1-psi}
    H \star \psi = E\psi\ .
\end{equation}
What about $W$? It turns out that the stationary Wigner function
is uniquely determined by the following two equations \cite{amer}
\begin{equation}\label{2-W}
    H \star W = W \star H = EW\ ,
\end{equation}
that is, $W$ satisfies the same equations as $\psi$ and
additionally it fulfills $W \star H = EW$ which is equivalent to
\begin{eqnarray}
 \left[\frac{1}{2m}\left(p +
\frac{i\hbar}{2}\partial_q\right)^2
+V\left(q-\frac{i\hbar}{2}\partial_p\right)\right]W(q,p)  \nonumber \\
&& \hspace*{-4cm}  =EW(q,p)\ .
\end{eqnarray}
We stressed, that there are infinite solutions of (\ref{1-psi})
and there is only one solution of (\ref{2-W}). Clearly, the Wigner
function does belong to the solutions of (\ref{1-psi}). Indeed,
taking
\begin{equation}\label{gW}
    g(y) = \varphi^*(-y/2)\ ,
\end{equation}
the formula (\ref{x1}) implies
\begin{eqnarray}  \psi(\Gamma) = 
\int dy\ e^{-{ipy}/{\hbar}}
 \varphi^*\left(q-\frac{y}{2}\right)
\varphi\left(q+\frac{y}{2}\right) ,
\end{eqnarray}
that is, $\psi(\Gamma)= 2\pi\hbar W$, where $W(q,p)$ is a Wigner
function corresponding to $\varphi$. Note, that $g(y)$ defined via
(\ref{gW}) satisfies (\ref{g2-s}) and hence the corresponding
kernel does indeed satisfy (\ref{KK}).

 As
an example let us compare the solutions of (\ref{1-psi}) and
(\ref{2-W}) for the harmonic oscillator. Taking $g(y)$ as in
(\ref{g-ex}) one obtains from (\ref{x1})
\cite{Torres1,har,li,scripta} the following formulae corresponding
to $n$th energy eigenstate $\varphi_n(q)$:
\begin{equation}\label{}
\psi_n(\Gamma) = \frac{1}{\sqrt{n!}}\, \Gamma^{* n}\, \exp\left( -
\frac{H}{2\hbar\omega}\right)\ ,
\end{equation}
with $\Gamma$ given by (\ref{Gamma}), whereas
\begin{eqnarray}\label{W-n}
W_n(q,p) =
\frac{(-1)^n}{\pi\hbar}L_n\left(\frac{4H}{\hbar\omega}\right)
\exp\left(-\frac{2H}{\hbar\omega}\right)\ ,
\end{eqnarray}
where $L_n$ denotes $n$th  Laguerre polynomial. Due to $|\Gamma|^2
= H/\hbar\omega$, one has for the probability distribution of
transition from $\varphi_n$ to the coherent state $|\Gamma
\rangle$
\begin{equation}\label{}
    |\psi_n(\Gamma)|^2 =  \frac{1}{{ n!}}\,
    \left(\frac{H}{\hbar\omega}\right)^n \exp\left( - \frac{H}{\hbar\omega}\right)\
    .
\end{equation}
Clearly, both $|\psi_n|^2$ and $W_n$ depends only upon the
oscillator energy $H$ and both are normalized according to
\begin{equation}\label{}
    \int |\psi_n|^2\, d\Gamma = \int W_n\, dqdp = 1\ .
\end{equation}
Moreover, it its easy to show
\begin{equation}\label{W2}
\int W^2_n\, dqdp = \frac{1}{2\pi\hbar}\ .
\end{equation}

Now, the family of Wigner-Weyl transforms ${\cal F}^s_{\rm WW}$
enables one to introduce the following family of Wigner functions:
$W_s = ({\cal F}^s_{\rm WW})^{-1}(\hat{\rho})$, where $\hat{\rho}$
stands for the density operator. One finds
\begin{eqnarray}\label{}
   W_s(q,p) &=& \frac{1}{2\pi\hbar}\int
   dy\,
e^{{i}py/{\hbar}}\,\nonumber \\ &\times & \Big\langle
q-(1-s)\frac{y}{2}\Big|\,\hat{\rho}\, \Big|\,
q+(1+s)\frac{y}{2}\Big\rangle ,
\end{eqnarray}
which reduces to $W(q,p)$ for $s=0$. The family $W_s$ was
introduced by Cahill, Glauber, Agarwal and Wolf
\cite{Cahill,Agarwal}. It satisfies two basic properties: it is
normalized
\[   \int W_s(q,p)\, dqdp = 1\ , \]
and for any quantum observable $\hat{F}$
\[   {\mbox Tr}(\hat{F}\hat{\rho}) = \int W_s(q,p) f_{-s}(q,p)
dqdp\ , \] where $ f_{-s} = ({\cal F}_{\rm
WW}^{-s})^{-1}(\hat{F})$. For $s=0$ the last formula reproduces
well known property of the Wigner function
\[  {\mbox Tr}(\hat{F}\hat{\rho}) = \int W(q,p) f(q,p)
dqdp\ . \] Moreover, $W_s$ provides correct quantum marginals:
\begin{eqnarray}\nonumber
\int dq\ W_s(q,p)&=&\langle p\vert\hat{\rho}\vert p \rangle\\
\label{x14} \int\ dp\ W_s(q,p)&=&\langle q\vert\hat{\rho}\vert q
\rangle\ .
\end{eqnarray}

It turns out that stationary $s$-Wigner functions $W_s$ are
uniquely determined by
\begin{equation}\label{2-W-s}
    H \star_s\, W_s = W_s \star_s\, H = EW_s\ .
\end{equation}
Equation (\ref{star-psi-s}) for $\psi_s$ has infinite number of
solutions whereas the set of two equations (\ref{2-W-s}) has only
one solution
\begin{eqnarray}   \label{W-s-phi}
\psi_s(\Gamma) &=&\int dy\ e^{-{i}py/{\hbar}}\nonumber
\\   &\times& \varphi^*\left(q-\frac{s+1}{2}y\right) \label{x15}
\varphi\left(q+\frac{1-s}{2}y\right) ,
\end{eqnarray}
i.e. $\psi_s(\Gamma)=2\pi\hbar W_s(q,p)$. Therefore, $W_s$ is only
one particular solution of (\ref{star-psi-s}). It is easy to see
that taking
\begin{equation}\label{}
    g_s(y) =  \varphi^*\left(-\frac{s+1}{2}y\right)
\end{equation}
in (\ref{x9}) one obtains $\psi_s(\Gamma)$ given by
(\ref{W-s-phi}). In particular for $s=1$ one obtains so called
Kirkwood--Rihaczek function $K(q,p)=W_{s=1}(q,p)$ which in the
case of pure state $\varphi$ reduces to
\begin{equation}\label{}
    K(q,p) = e^{ipq/\hbar}\, \widetilde{\varphi}^*(p)\varphi(q)\ .
\end{equation}
It was introduced by Kirkwood \cite{kirk} as an alternative for
the  Wigner  function. Then, in 1968, the same formula was
rediscovered by Rihaczek \cite{Rihaczek} in the context of signal
time--frequency distributions (see \cite{Englert} for a useful
review). Recently, this function was analyzed and applied in
various contexts  in \cite{wod-p1,wod-p2,tomK,proK,therK,Zak}.

\section{DISCUSSION}\label{A5}

Both the phase space wave function $\psi(\Gamma)$ and s-ordered
Wigner function $W_s$ encode the entire information about the
quantum state $\varphi(q)$. Due to the basic property
\begin{equation}\label{prob}
    \int |\psi(\Gamma)|^2\, d\Gamma = 1\ ,
\end{equation}
some authors call $|\psi(\Gamma)|^2$ a probability distribution in
phase space. Clearly, quantum mechanics does not allow for a
genuine probability distribution in $q$ and $p$! To interpret
$|\psi(\Gamma)|^2$ correctly note that formula (\ref{Kernel}) may
be rewritten as the following inner product
\begin{equation}\label{}
    \psi(\Gamma) = \langle \varphi_\Gamma|\varphi\rangle\ ,
\end{equation}
where
\begin{equation}\label{}
    \varphi_\Gamma(q') = K^*(\Gamma;q')\ .
\end{equation}
Let us consider kernels defined by (\ref{K-g}). Then,  due to
(\ref{g2}), $\varphi_\Gamma(q')$ is a normalized wave function in
the position representation. Therefore, $|\psi(\Gamma)|^2$ is the
probability density of transition from the state $\varphi$ to
state $\varphi_\Gamma$. In particular for the coherent state
kernel $K_{\rm CS}$ one has $|\psi(\Gamma)|^2 =
|\langle\Gamma|\varphi\rangle|^2$ which defines the Husimi
function for the state $\varphi$.

Now, the Wigner function defined quasi-distribution such that
$\int W\,dqdp=1$. Since $W$ is a special solution of the
Schr\"odinger equation in phase space one has
\begin{equation}\label{}
    W_\varphi(q,p) = \frac{1}{2\pi\hbar}\, \langle \varphi_\Gamma|\varphi\rangle\ ,
\end{equation}
where $W_\varphi$ is the Wigner function corresponding to
$\varphi$. Formulae (\ref{Ksg}) and (\ref{gW}) imply
\begin{equation}\label{}
    \varphi_\Gamma(q') = 2 \varphi^*(2q-q')\, e^{-2ip(q'-q)/\hbar}\ .
\end{equation}
 It should be stressed that
the phase formulation based on the wave function $\psi(\Gamma)$ is
restricted to pure states only whereas the approach based on
Wigner function works perfectly for general mixed states $\rho$.
Therefore, this approach is much more general. Note, that for
mixed states one has
\begin{equation}\label{}
    \int W^2\, d\Gamma \leq \frac{1}{(2\pi\hbar)^2}\ ,
\end{equation}
and the equality holds for pure states only. Therefore, for
general mixed states $(2\pi\hbar)^2\, W^2$ cannot be interpreted
as a probability distribution.

\acknowledgments

This work was partially supported by the Polish State Committee
for Scientific Research Grant {\em Informatyka i in\.zynieria
kwantowa} No PBZ-Min-008/P03/03.

 \end{document}